\documentstyle[12pt]{article}

\begin{document}

\newcommand{\dslash}{\mbox{ $\not \!\! D$}}

\newcommand{\unmezzo} {\frac{1}{2}}
\newcommand{\unterzo} {\frac{1}{3}}
\newcommand{\unquarto}{\frac{1}{4}}

\newcommand{\parzialet}[1]   {  \frac{ \partial{#1} }{\partial{t}} }
\newcommand{\parzialexi}[1]  {  \frac{ \partial{#1} }{\partial{x_i}} }
\newcommand{\parzialexxi}[1] {  \frac{ {\partial}^2{#1} }{\partial {x_i^2}} }
\newcommand{\derivatax}[1] {  \frac{        d{#1} }{        dx} }
\newcommand{\derivatat}[1] {  \frac{        d{#1} }{        dt} }

\newcommand{\traccia}{ {\rm Tr} }

\newcommand{\be}{\begin{equation}}
\newcommand{\ee}{\end{equation}}
\newcommand{\ca}{ {\cal A} }
\newcommand{\ch}{ {\cal H} }

\title{Non-Exponential Relaxation Time Scales
in Disordered Systems: an Application to Protein Dynamics}

\author{Giulia IORI$^{(*)}$, Enzo MARINARI$^{(*)}$ and Giorgio PARISI \\[1.5em]
Dipartimento di Fisica,\\
Universit\`a di Roma {\it Tor Vergata},\\
Viale della Ricerca Scientifica, 00173 Roma, Italy\\
and\\
Infn, Sezione di Roma {\it Tor Vergata}\\
{\footnotesize iori@roma2.infn.it marinari@roma2.infn.it
        parisi@roma2.infn.it}\\[1.0em]
{\small $(*)$:  and Dept. of Physics and Npac,} \\
{\small Syracuse University}\\
{\small Syracuse, NY 13244, USA}}
\maketitle

\vfill
\begin{flushright}
  {\bf ROM2F-92-35}\\
  {\bf SCCS 331}\\
  {\bf hep-lat/9208001 }\\
\end{flushright}

\vfill
\newpage

\begin{abstract}

We study the dynamics of an heteropolymeric chain relaxing toward a
new equilibrium configuration after the action of an external
perturbation.  We compare the results from Monte Carlo simulations
with the results of a Langevin normal mode decomposition.  We discuss,
for sake of comparison, the case of an ordered homopolymeric chain.

\end{abstract}

\vfill

\newpage

\section{Disordered System Dynamics}

Disordered system$^{\cite{BOOK1,BOOK2}}$
potential relevance for understanding protein folding has
received much attention in the last few years (see for example refs.
\cite{BRYWOL,GARORL,SHAGUT,MEZPAR,FRAU}.
They differ from their ordered
counterpart in being characterized from a dynamical behavior which is
different on different (and many) time scales.
A large number of experiments (for a discussion see for example
ref.~\cite{BOOK2}) has been done in order to probe such complex
dynamical features (where up to $15$ different orders of magnitude can
be relevant).

Typically a single experiment can only probe one scale at time, since
the relaxation processes which occur on faster time scales can be
already equilibrated when the ones with larger characteristic times
are yet completely frozen.  It is fortunate that in the {\em aging}
experiments$~{\cite{AGING}}$ the relevant degrees of freedom act on at
least two time scales simultaneously.

These dynamic phenomena are called {\em dispersive} as they are
characterized by long tailed waiting time distribution where the first
moment diverges. Many toy models (see for example ref. \cite{SHLKLA})
have been proposed to shed light on such multi-timescale systems.

These models are generally explicitly hierarchical.  It would be more
satisfying to have model in which a hierarchical structure would
emerge naturally, as for example happens in the Sherrington
Kirkpatrick model (SK).

The description of a very simple model$^{\cite{BOOK2}}$ can be useful.
It describes a particle at temperature $T$ which can hop among
hierarchically grouped states. One can study the solution of the
Fokker-Planck equation for the probability $P(x,t)$ of finding the
particle in a given one of the allowed states.  One introduces a
matrix of transition rates between states that has the same kind of
ultrametrix structure as the $q$ replica order parameter matrix in the
mean field theory.  A possible interesting choice for the
dependence of the transition rates on the ultrametric distance is
assuming that the barriers $\delta$ between states increase linearly
(with a proportionality constant $\Delta$) with the ultrametric
distance $k$

\begin{equation}
  \delta(k)= \Delta \ k \ .
\end{equation}

In this case one can find that the probability $P_0$ (which in a spin
system can be identified with the spin autocorrelation function) of
remaining in the initial state evolves according to a power law

\begin{equation}
   P_0 \simeq t^{\frac{-T \log 2}{\Delta}} \ .
\end{equation}

If the barriers grow too slowly with the ultrametric distance the
system delocalizes. The case which is most interesting for us is the
{\em marginal} one, where we select for the barrier height $\delta(k)$ a
logarithmic dependence over the ultrametric distance $k$:

\begin{equation}
   \delta(k) = \Delta \ \log k\ .
\end{equation}

In that case the time evolution of $P_0$ is found to have a {\em
stretched exponential} dependence (as first introduced by
Kohlrausch$^{\cite{KOL}}$):

\begin{equation}
  P_0 \simeq \exp(-t^{ \frac {T} {\Delta}})\ .
\end{equation}

In the following we will focus our attention to such stretched
exponential behaviors, as found in the dynamics of the heteropolymeric
chains.

In Section \ref{SECSG} we discuss what happens in disordered spin
systems. In Section \ref{SECNED} we define the quantities needed to
discuss relaxation dynamics. In Section \ref{SECMC} we recall the
definition of our model (IMP in the following, see ref. \cite{IMP}.
For further developments on this model see refs.
\cite{FUKU2,MVS,PLIMAR}),
we define our Monte Carlo procedure, and we present the results of
numerical simulations.  In Section \ref{SECLE} we introduce the
Langevin Equation approach and its connection to a stretched
exponential dynamics, and we discuss our results from a normal mode
decomposition.

\section{Spin Glass Dynamics on Large Time Scales
\protect\label{SECSG} }

We will be interested in studying time-dependent effects in the broken
phase of a disordered system. The typical example is the replica
broken phase of a mean-field SK model$^{\cite{BOOK1}}$.  In this case
broken ergodicity makes the analysis very difficult: one has to be
careful in examining phenomena typical of the different time scales
relevant for the dynamical evolution.

Let us assume that we are dealing with a system which has many pure
states. In this case for a sample in the thermodynamic limit (infinite
volume) the time evolution is confined to one of the available pure
states: the system will not be able to escape to another valley in a
finite time.  This relaxational dynamics inside one state has been
studied in mean field theory by Sompolinsky and
Zippelius$^{\cite{SOMZIP}}$ and shows a power law time decay
(for the case of an IMP heteropolymer a power behavior for the
dynamics inside one state has been found in the numerical simulations
of ref. \cite{PLIMAR}).

Many efforts have also been done to understand the dynamics on time
scales which diverge with the volume of the system. In such a way one
can focus on time scales which, for increasing volume, allow
transitions between the different states (indeed the precursors of
what will be the states in the thermodynamical limit).  The basic
hypothesis, put forward by Sompolinsky$^{\cite{SOMPOL}}$, is here that
the large time limit of the dynamic evolution is governed by a strong
hierarchy of time scales.

The precise understanding of the asymptotic regime require a detailed
knowledge of the properties of the system for large finite size: for
example informations about the height of the free energy barriers
between states can give lot of helpful information.  One reasonable
assumption, that one needs in absence of more detailed informations
about the barriers between the states, consists in assuming that the
time needed for the system to jump from one state to another is
correlated with the distance between the two states. Under this
assumption one can associate to each value of the distance $k$ a
typical hopping time scale $\tau$.  At any given time $\tau$ the
system is described by a given equilibrium measure.

Mackenzie and Young$^{\cite{MACYOU}}$ have studied numerically the
maximum relaxation time $\tau_M$ as a function of the system size $N$,
for the mean field SK system.  They have found that

\begin{equation}
  \log(\tau_M) \simeq N^{\frac{1}{4}} \ .
\end{equation}

Similar studies on equilibrium relaxation dynamic have also been done
on system like the Random Energy Model$^{\cite{REM}}$ where no
hierarchical organization between states is present (since replica
symmetry is broken only at first order).  Work by De Dominicis, Orland
and Lain\'ee$^{\cite{DEDETA}}$ shows that also this model exhibit a
non-exponential relaxation, suggesting that ultrametricity is not
necessarily needed to obtain this kind of behavior. A stretched
exponential dynamical behavior has also been find in the case of the
Spherical Spin model in ref. \cite{JAG}.

\section{Non Equilibrium Dynamics \protect\label{SECNED} }

In the following we will look at the problem of the relaxation
dynamics of a disordered system not far from equilibrium. The system,
at equilibrium for some given values of the external parameters,
undergoes an external perturbation (some of the parameter values are
abruptly modified). We will study the subsequent search of a new
equilibrium configuration.

Let us assume that
the system just goes out of its local (free energy) minimum, without
moving too far away from its initial conformation.
Eventually we expect the system to relax in the free energy valley of
the chosen minimum. Here the system will only undergo very small
conformational modifications. We will show this relaxation to have a
stretched exponential form.

This matter has been studied experimentally in the pioneering work on proteins
by Frauenfelder and collaborators (see ref. \cite{FRA} and references therein).
They measure different quantities $\ca(t,T)$, which depend on the time $t$
and on the temperature $T$, and define the relaxation functions as

\be
  \Phi(t) \equiv \frac{\ca(t,T) - \ca(\infty,T)}{\ca(0,T)-\ca(\infty,T)}\ ,
  \protect\label{F_COR}
\ee
where the time $t=0$ is the one just after the perturbation was applied.
After a stress in temperature or pressure $\Phi(t)$ was shown to have a
non-exponential behavior.

In the next sections of this note we will study the dynamical behavior
of an heteropolymeric system. We will look at what we believe to be
the relaxation in a given free energy minimum (see the former
discussion), but we will be working at times which are very shorter
than the experimental ones (which can be of the order of seconds). We
will show that in this case we cannot explain our results with a
simple exponential relaxation law, and that the introduction of a
stretching parameter will lead us to a very good fit of the numerical
results.

An analytical study of the Langevin equation normal modes decomposition will
show how such temporal evolution can be associated to the existence of a
long tail (in the small frequency region) in the eigenvalue spectrum of the
relevant dynamical operator.

\section{Montecarlo Simulations
\protect\label{SECMC} }

Let us start by recalling the definition of the IMP
model$^{\cite{IMP}}$. We consider a chain formed of $N$ sites (they
would be identified, in the protein analogy,with {\em sequences} of
amino-acids, i.e. preassembled segments of the secondary structure):
their position in {\em continuum} $3$ dimensional space is
characterized by the $3$ values of the coordinates $x_i^\mu$.  We
define the energy between two sites of the chain as

\begin{equation}
  E_{i,j} \equiv
  \delta_{i,j+1} \
  r_{i,j}^2
  + \frac{R}{r_{i,j}^{12}}
  - \frac{A}{r_{i,j}^{6}}
  + \frac{\eta_{i,j}}{r_{i,j}^{6}} \ ,
  \protect\label{F_ENE}
\end{equation}
where $r_{i,j}$ is the usual Euclidean distance between the sites $i$ and $j$.

The harmonic term couples the first  neighbors on the chain. The deterministic
part of the potential has the  usual Lennard-Jones form. The quenched
disordered part of the potential is built on the $\eta$ variables, which are
distributed with a zero expectation value and with a second moment

\begin{equation}
  \langle \eta_{i,j} \ \eta_{k,l} \rangle = \epsilon \ \delta_{(i,j),(k,l)}\ ,
\end{equation}
where $\langle \rangle$ indicates the expectation value over the quenched
distribution, which we have chosen to be uniform, $\epsilon$ characterizes the
disorder strength, and $\eta_{i,j} = \eta_{j,i}$.

The Hamiltonian of the model is defined as

\begin{equation}
  H \equiv \sum_{i=1}^{N} \sum_{j>i} E_{i,j}\ .
      \protect\label{HAMI}
\end{equation}

In our typical run we have selected values of the parameters close to the ones
of ref. \cite{IMP}. The system is, indeed, in what in ref. \cite{IMP} we have
recognized as the {\em folded} phase of the heteropolymeric chain. We have
studied chains done of $N=15$ and $30$ sites, for $R=2.0$, $A=3.8$ and
$\epsilon=6$.

We have tried to build a numerical experiment close to the
experimental conditions of the Frauenfelder group true
experiments$^{\cite{FRA}}$ (also if, as we said, our time scales are
very shorter). We have been starting from a chain thermalized at
$\beta=1$, and we have abruptly decreased its temperature $T$
($T\equiv \frac{1}{\beta}$) to values ranging from $\beta=2$ to
$\beta=10$. We have computed the relaxation function (\ref{F_COR}) by
choosing as observables $\ca$ the internal energy, the gyration radius
of the chain and the link length (for detailed definitions see ref.
\cite{IMP}). For each $\beta$ value we have averaged the relaxation
functions over $500$ Monte Carlo stories.

We have been careful to check that the perturbation was small enough
not to produce big changes in the conformation of heteropolymer. The
system was only allowed to change in similar {\em quasi-states} (in
Frauenfelder terminology) and not to have a transition to a completely
different state. This is for us good evidence (consistently with
Frauenfelder experiments) that the non-exponential behavior we are
discussing is not generated from the visiting of comformationally
different minima with different underlying time scales, but from the
fact that the dynamics in a single minimum is itself non-exponential.

All our runs are consistent with a stretched exponential relaxation for the
correlation function with a stretching parameter $\gamma$ in the range of
order~$.5$.

In fig. 1 we show the energy decay, as a function of the Monte Carlo
time (in logarithmic scale), for four different values of the final
$\beta$, i.e. for stronger and weaker perturbations.

In Fig. 2a we show, as an example, the decay of the energy for
$\beta=2.0$ (after perturbing a configuration thermalized at
$\beta=1.0$) in a system done of $N=15$ sites. We get a very good
stretched exponential fit of the form

\be
  e^{-(\frac{t}{\tau})^\gamma} \ ,
  \protect\label{F_STR}
\ee
with $\gamma \simeq .54$. A fit with a pure exponential behavior would
not match the data. In Fig. 2b we show the same dependence for
$\beta=3.0$. Here $\gamma$ is smaller, $\sim .38$. For higher values
of $\beta$ (i.e. for stronger perturbations) we did not succeed to get
good fits. In this region the validity of the linear approximation is
far from being clear.

\section{The Normal Mode Analysis \protect\label{SECLE}}

The fluctuation-dissipation theorem shows that a simple relation between the
equilibrium fluctuations of a system and its linear response to an external
perturbation exists.

Let us consider the case where a constant field $h=\tilde{h}$ is
applied to the system for a long time. Eventually the system reaches
equilibrium. At a time, that we define as $t=0$, $h$ is switched off.  The
change in $h$, $\delta h = - \tilde{h}$, changes the average value of
the physical quantities from their original values at equilibrium with
$h=\tilde{h}$. If the field is weak we can assume that the change is a
linear functional of the field

\be
  \langle \ca (t)\rangle_h - \langle \ca \rangle_0
  = - \alpha(t) \ \tilde{h}\ ,
\ee
where $\alpha(t)$ is the relaxation function.

In the case the have discussed in the previous section the
perturbation field can be identified with the temperature variation

\be
  \tilde{h} = T_2 - T_1\ ,
\ee
where $T_1$ is the initial equilibrium temperature ($\beta=1$ in our
case), and $T_2$ is the final one. Under these conditions the
fluctuation-dissipation theorem gives

\be
  \alpha(t) = \frac{1}{k_B T}
  \langle \ca(t) \ca(0) \rangle_c \ ,
\ee
where the correlation function on the right hand side is the connected part.

In the following we will discuss how to compute time dependent
correlation functions $\langle \ca(t) \ca(0) \rangle_c$ (where $h$ has
been switched off at $t=0$). In order to do that we define the time
dependent non-equilibrium probability distribution function
$P(x,t)$, and discuss it in the framework of the Langevin equation.

The Langevin equation for the dynamics of a generic site of the
chain can be written as

\be
  \parzialet {x_i} = - \zeta \ \parzialexi {H(x)} + f_i \ .
  \protect\label{F_LAN}
\ee

The probability distribution of the solutions of eq. (\ref{F_LAN})
satisfies the Fokker-Planck equation

\be
  \parzialet{P(t)} = \sum_i
  \{
  - \frac{\partial}{\partial x_i} (\frac{F_i}{\zeta} P(t))
    + \Omega \parzialexxi{P(t)}
  \} \ ,
\ee
where $F_i$ is the force acting on the $i$-particle, and $\Omega$ is the
variance of the $f_i$.  In the following we will focus on oscillations
close to a minimum of the energy, so we will ignore the effect of the
thermal noise $f_i$, by setting $\Omega=0$, and we will use the fact that
the gradient of $H$ is zero. This means that, in the subsequent
analysis of numerical data, we will look at configurations at $T=0$
(cooled by steepest descent procedure).
The probability distribution function can be written as

\be
  P(x,t) = \psi_0(x) \rho(x,t)\ ,
\ee
where

\be
  \psi_0(x) \simeq e^{- \frac{\beta H}{2} } \ ,
\ee
is the eigenvector of $\ch_{FP}$ with zero eigenvalue, and in the
linearized approximation $\rho$ satisfies

\be
  \parzialet{\rho(x,t)} = - \ch_{FP}\  \rho \ ,
\ee
where

\be
  \ch_{FP} \equiv - \frac{1}{2} \parzialexxi{H}\ .
\ee

We define the orthonormal basis of eigenvectors $\{\psi_n\}$ of $\ch_{FP}$

\be
  \ch_{FP} \ \psi_n = \lambda_n \psi_n \ .
\ee

We can decompose the time dependent probability distribution as

\be
  P(x,t) = \psi_0(x) \sum_n c_n \psi_n(x) e^{-\lambda_n t}\ ,
\ee
where the $c_n$ are constant coefficients (which depend only on the initial
condition) and $c_0=1$.

The evaluation of the time dependent correlation function gives as result

\be
  \langle \ca(t)\ca(0) \rangle_c =
  \sum_{i=1}^{3 N} e^{-\lambda_i t}
  |< \psi_0 | \ca | \psi_i >|^2 \ .
  \protect\label{F_EIG}
\ee

We will use this relation to reconstruct the energy time dependent
correlation functions from the knowledge of the eigenvalues of the
Fokker-Planck Hamiltonian $\ch_{FP}$. The correlation function has
been expressed in (\ref{F_EIG}) as a sum of time exponential
functions, with decay factors

\be
  \tau_i \equiv \frac{1}{\lambda_i}\ .
\ee

Such an sum can (and will indeed) generate a stretched
exponential behavior. This can happen if the smallest $\lambda_i$ is not very
different from the larger ones,  i.e. if there is no large {\em gap}.
We expect that to happen indeed it is well known, for simple
disordered systems (for example a disordered chain in $d=1$, see ref.
\cite{DYSON}), that the frequency spectrum of the normal modes has a
sizeable tail (in the small frequency region), which is absent in the
corresponding ordered system. Such a difference from the ordered case
tends to become more important with increasing space dimensionality.

We have looked at chains of $N=15$ and $N=30$ sites, by comparing the
homopolymeric chain ($\epsilon=0$) with the strongly disordered
heteropolymer ($\epsilon=6$). We have computed the $\ch_{FP}$ eigenvalues
for the minimum energy states (at $T=0$) by using an imsl routine.

We have obtained a stretched exponential behavior for times up to
$\sim 100$ in equation (\ref{F_EIG}). In the direct evaluation of the
correlation function of section (\ref{SECMC}) we have found it holds
up to $\sim 10000$ Monte Carlo chain sweeps. This is very smaller than
the macroscopic experimental time scales (corresponding to $\sim
10^{10}$ Monte Carlo chain sweeps).

In Fig. 3a we plot the reconstructed correlation functions, for
$\epsilon=6$ and for the pure homopolymer $\epsilon=0$, for $N=15$
sites.  The stretched exponential fit (with $\gamma=.33$) works very
well in the disordered case, while the best fit (with $\gamma=.56$) is
very inadequate for the homopolymer. In Fig. 3b the same fits for
$N=30$, where we find a very similar qualitative behavior: here
$\gamma=.39$ for $\epsilon=6$, and the $\epsilon=0$ fit, with
$\gamma=.58$, is very bad. For this region of time we have been able
to exclude a simple pure exponential behavior in the {\em folded}
phase.  For long times, as expected, the behavior becomes exponential.
We have checked that an exponential fit gives, in this region, a very
good result, with a decay which is characterized from the lowest
eigenvalue of $\ch_{FP}$.

\section*{Acknowledgments}

We thank Luca Biferale, Pawel Pliszka and Felix Ritort for interesting
conversations, and especially Maria Vittoria Struglia for continuous
discussions and help, and for providing us with the $T=0$ chain
configurations we used to compute normal modes.

\vfill
\newpage
%___________________________________________________________________
\section*{Figure Captions}
\begin{itemize}

  \item{Fig. 1. }

     Energy time decay (eq. (\ref{F_COR})) versus Monte Carlo time for a
chain done of $N=15$ points, relaxing from $\beta=1.0$ to
$\beta = 2.0, 3.0, 8.0, 10.0$. The point labelled with $\beta=1$ is
the starting value of the energy for the four different relaxation runs.

  \item{Fig. 2a. }

     Energy time decay (eq. (\ref{F_COR})) versus Monte Carlo time for a
chain done of $N=15$ points, relaxing from $\beta=1.0$ to
$\beta = 2.0$.
The open dots are from the Monte Carlo data (average over
$500$ time periods), the smooth curve is the best, stretched
exponential fit, with a stretching exponent $\gamma = .54$

  \item{Fig. 2b. }

     As in fig. 2a, but final $\beta = 3.0$. Here $\gamma = .38$.

  \item{Fig. 3a. }

    The reconstructed energy connected correlation functions, for
$\epsilon=6$ and for the pure homopolymer $\epsilon=0$, for $N=15$
sites.

  \item{Fig. 3b. }

    As in Fig. 3a, but $N=30$.

\end{itemize}
\vfill
%___________________________________________________________________________

\newpage

\begin{thebibliography}{99}
  \bibitem{BOOK1}
    M. Mezard, G. Parisi and M. Virasoro,
    {\em Spin Glass Theory and Beyond}
    (World Scientific, Singapore 1987).
  \bibitem{BOOK2}
    K. H. Fischer and J. A. Hertz,
    {\em Spin Glasses}
    (Cambridge University Press, Cambridge, UK 1991).
  \bibitem{BRYWOL}
    J.~D.~Bryngelson and P.~G.~Wolynes,
    Proc. Natl. Acad. Sci. USA {\bf 84} (1987) 7524.
  \bibitem{GARORL}
    T.~Garel and H.~Orland,
    Europhys. Lett. {\bf 6} (1988) 307.
  \bibitem{SHAGUT}
    E.~I.~Shakhnovich and A.~M.~Gutin,
    Europhys. Lett. {\bf 8} (1989) 327.
  \bibitem{MEZPAR}
    M.~Mezard and G.~Parisi, J. Physique {\bf I1} (1991) 809.
  \bibitem{FRAU}
    I. Iben, D. Braunstein, W. Doster, H. Frauenfelder, M.~K.~Hong,
    J.~B.~Johnson, S. Luck, P.~Ormos, A. Schulte, P.~J.~Steinbach,
    A.~H.~Xie and R.~D.~Young,
    Phys. Rev. Lett. { \bf 62} (1989) 1916.
  \bibitem{AGING}
    L.~Lundgren, P. Svedlindh, P. Nordblad and O. Beckman,
    Phys. Rev. Lett. {\bf 51} (1983) 911;
    P. Nordblad, P. Svedlindh, L.~Lundgren,  and L.~Sandlund,
    Phys. Rev. {\bf B33} (1986) 645.
    Ph.~Refregier et al., J. Phys. (Paris) {\bf 48} (1987) 1533;
  \bibitem{SHLKLA}
    M.~F.~Shlesinger and J.~Klafter,
    {\em The Nature of Temporal Hierarchies Underlying Relaxation in
    Disordered Systems,}
    in {\em Fractals in Physics,}
    edited by L. Pietronero and E. Tosatti (Elsevier, Amsterdam 1986).
  \bibitem{KOL}
    F.~Kohlrausch, Pogg. Ann. Phys. {\bf 119} (1863) 352.
  \bibitem{IMP}
    G. Iori, E. Marinari and G. Parisi,
    J. Phys. {\bf A} (Math. Gen.) {\bf 24} (1992) 5349.
  \bibitem{FUKU2}
    M. Fukugita, D. Lancaster and M. G. Mitchard,
    J. Phys. Lett. {\bf A} (Math. Gen.) {\bf 25} (1992) L121.
  \bibitem{MVS}
    M. V. Struglia, {\em Conformational Properties of Random Heteropolymers
    in the Folded Phase}, Roma {\em Tor Vergata} preprint ROM2F/92/42
    (1992).
  \bibitem{PLIMAR}
    P. Pliszka and E. Marinari
    {\em On Heteropolymer Shape Dynamics}
    Syracuse preprint SCCS 330, hep-lat 9207011 (Syracuse, July 1992).
  \bibitem{SOMZIP}
    H. Sompolinsky and A. Zippelius, Phys. Rev. {\bf B25} (1982) 6860.
  \bibitem{SOMPOL}
    H. Sompolinsky, Phys. Rev. Lett. {\bf 47} (1981) 359.
  \bibitem{MACYOU}
    N. D. Mackenzie and A. P. Young, Phys. Rev. Lett. {\bf 49} (1982) 301.
  \bibitem{REM}
    B. Derrida, Phys. Rev. {\bf B24} (1981) 2613.
  \bibitem{DEDETA}
    C. de Dominicis, H. Orland and F. Lain\'ee, J. Physique {\bf 46} (1985)
    L463.
  \bibitem{JAG}
    A. Jagannathan in {\em Biologically Inspired Physics,} edited by L. Peliti
    (Plenum Press, New York, USA 1991), p. 365.
  \bibitem{FRA}
    H. Frauenfelder, K. Chu and R. Philipp, in {\em Biologically Inspired
    Physics,} edited by L. Peliti (Plenum Press, New York, USA 1991), p. 1.
  \bibitem{DYSON} F. J. Dyson, Phys. Rev. {\bf 91} (1953) 1331.
%__________________________________________________________________
\end{thebibliography}
\end{document}